\documentclass[aps,prl,twocolumn,groupedaddress,showpacs]{revtex4}

\usepackage{graphicx}

\def\prb{Phys. Rev. B }
\def\prl{Phys. Rev. Lett. }

\def\be{\begin{equation}}
\def\ee{\end{equation}}
\def\ba{\begin{eqnarray}}
\def\ea{\end{eqnarray}}

\begin{document}

\title{Correlated tunneling and the instability of the fractional quantum Hall edge }

\author{Dror Orgad and Oded Agam}
\affiliation{The Racah Institute of Physics, The Hebrew University,
Jerusalem, 91904, Israel}

\date{\today}

\begin{abstract}
We consider a class of interaction terms that describes correlated
tunneling of composite fermions between effective Landau levels.
Despite being generic and of similar strength to that of the usual
density-density couplings, these terms are not included in the
accepted theory of the edges of fractional quantum Hall systems.
Here we show that they may lead to an instability of the edge towards
a new reconstructed state with additional channels, and thereby demonstrate
the incompleteness of the traditional edge theory.
\end{abstract}

\pacs{71.10.Pm}

\maketitle

Much of our theoretical understanding of the edge physics in the fractional
quantum Hall (FQH) regime is based on the chiral Tomonaga-Luttinger (CTL) model
put forward by Wen \cite{Wen-papers}. For the Jain fractions $\nu=n/(2n \pm 1)$,
this low-energy effective theory may be derived by considering small fluctuations
around the mean-field configuration which describes the bulk of the sample as
$n$ filled Landau-levels of composite fermions \cite{Wen-papers,cslut,compedge}.
The theory associates a chiral channel with each of these levels where it crosses
the Fermi energy near the edge, and includes the electronic correlations of the
bulk FQH liquid through a non-trivial exchange statistics between the excitations of
the different channels.

A central prediction of this edge model is a non-linear tunneling conductance
$I\sim V^\alpha$, with an exponent $\alpha=3$ over a range of filling factors
$1/3<\nu<1/2$. While tunneling experiments
\cite{Chang96,Grayson98,Chang01,Hilke01} have indeed found a power-law
characteristic, the exponent scales as $\alpha\approx a/\nu+b$, where $a$ and $b$
are non-universal constants, in contradiction to the predicted value \cite{Chang03Levitov01}.
This discrepancy suggests that the accepted CTL theory does not incorporate
some important elements for the understanding of the FQH edge physics.
In this Letter we concentrate on one such element, namely a class of interaction
terms describing correlated tunneling of composite fermions between edge states,
and show that it may lead to edge reconstruction, where the new structure supports
additional channels.

In order to clarify the origin of such processes, consider the
interaction energy between the electrons written in terms of composite fermions.
To this end we transform from the original
electronic operator $\Psi({\bf r})$ to a composite fermion operator
$\psi({\bf r})=e^{i\Lambda({\bf r})}\Psi({\bf r})$,
via a Chern-Simons phase $\Lambda({\bf r})$, and expand $\psi({\bf r})$
using the composite fermion annihilation operators on the $n$ occupied
effective Landau levels $\psi({\bf r})=\sum_{i=1}^n \psi_i({\bf r})$ \cite{comment1}.
The result is
\begin{eqnarray}
\label{interaction}
&&\!\!\!\frac{1}{2}\int d^2r d^2r' \rho_e({\bf r})V({\bf r}-{\bf r'})\rho_e({\bf r'}) \\
&&\!\!\!=\frac{1}{2} \sum_{ ijkl=1}^n \int d^2r d^2r'
\psi_i^\dagger({\bf r})
 \psi_j({\bf r}) V ({\bf r}\ - {\bf r'})\psi_k^\dagger({\bf r'})\psi_l({\bf r'}),
 \nonumber
\end{eqnarray}
where $\rho_e=\Psi^\dagger\Psi$. In the traditional CTL model only the Hartree terms
$i=j$, $k=l$ are kept. When the interaction potential is short ranged the Fock combination
$i=l$, $j=k$, gives a similar contribution but with an opposite sign and may be incorporated
via a renormalization of the density-density coupling. Our focus lies with the effect of
additional terms with $i=j$
but $k\neq l$ (or vice versa) in (\ref{interaction}). These processes, which we dub correlated
tunneling, correspond to tunneling of composite fermions between different Landau
levels with an amplitude that depends on the total density.
Correlated tunneling is of the same order as the Hartree-Fock interactions \cite{longpaper}
when the interaction strength is comparable to the gap between the effective Landau levels.
This condition, which is typically satisfied in the FQH regime,
ensures that correlated tunneling conserves both energy and momentum.

The possibility of edge reconstruction due to the competition between
a long-range Hartree term and a short-range exchange interaction, has
been realized in the context of the integer quantum Hall effect \cite{Chamon94}.
Composite fermions Hartree-Fock calculations have found a similar transition,
triggered by a softening of the confining potential, in a $\nu=1/3$ edge (but not
at higher fillings) \cite{Murthy03}. Edge reconstruction was also detected in exact
diagonalization studies of FQH systems with sharp edges
\cite{Tsiper01,Wan02,Wan03}, and was attributed to the finite range of the interactions
\cite{Yang03}. Here we offer a new avenue for the instability via the correlated
tunneling processes.

A qualitative picture for the nature of this instability follows from a toy model
where two states, occupied by
identical charged bosons, are coupled by correlated tunneling.
Let $N_1$ and $N_2$ represent the number of particles in
each one of these states, and $H_C=(N_1+N_2-N_T)^2/2C$ be the charging
energy of the system, where $C$ plays the role of capacitance, and $N_T$ is a
constant fixed by the chemical potential. The correlated tunneling
is represented by a term of the form
$H_{CT}=\lambda \left(N_1+N_2\right) \left(b_1^\dagger b_2+ b_2^\dagger
b_1 \right)$, where $b_j^\dagger$ denotes
a creation operators of a particle in state $j$, and $\lambda$
is a constant which characterizes the strength of the process.
The total Hamiltonian of the system is $H=H_C+H_{CT}$, and
for $\lambda=0$ its ground state
is set by the condition $N_1+N_2=N_T$. For $\lambda \neq 0$
the problem can be simplified by the transformation
$b_\pm=(b_1 \pm b_2)/\sqrt{2}$, which brings the Hamiltonian
to the form
\begin{equation}
H= \sum_{\alpha, \beta =\pm}
A_{\alpha \beta}N_\alpha N_\beta- \frac{N_T}{C}(N_+ + N_-) +\frac{N_T^2}{2C},
\label{eq:toy}
\end{equation}
where  $N_\pm= b_\pm^\dagger b_\pm$, and
\begin{eqnarray}
A=\frac{1}{2C}
 \left( \begin{array}{cc} 1+ 2C \lambda  & 1 \\
1  & 1-  2C \lambda \end{array} \right).
\end{eqnarray}
From Eq. (\ref{eq:toy}) it is evident that the eigenvalues of the
matrix $A$ determine the nature of the ground state. For any finite
$\lambda$, one of these eigenvalues, $(1 \pm
\sqrt{1+4\lambda^2 C^2})/2C$, is negative, and the system becomes
unstable. This instability drives $N_- \to \pm \infty$ while $N_+
\approx (1 -\lambda C) N_T- (1-2\lambda C)N_- $ for $\lambda C \ll
1$. Thus, unlike the case $\lambda=0$ where the minimum energy
occurs for zero charging, correlated tunneling leads to a divergence
of the total charge $N_1+N_2-N_T\simeq C \lambda(2N_- -N_T) \to \pm
\infty$. This is because depleting or filling up the system with
particles, reduces the kinetic energy associated with the tunneling.

This simple consideration points to the possibility that correlated tunneling may
also play an important role in determining the nature of the ground state
of the FQH edge. As a test case for such a scenario we focus
our attention on the $\nu=2/5$ edge and take as our starting point the
CTL model whose action in imaginary time is
\ba
\nonumber
S_0&=& \frac{1}{4\pi}\int dx d\tau \sum_{i,j=1}^2 \left(
i \partial_x \phi_i K^{-1}_{ij} \partial_\tau \phi_j + \partial_x
\phi_i V_{ij}  \partial_x \phi_j \right) \\
&+&\frac{\pi}{L}\int d\tau \sum_{i,j=1}^2 N_i V_{ij} N_j .
\label{S0}
\ea
By inverting the Chern-Simons transformation, the electronic operator may be expanded
according to $\Psi({\bf r})=e^{-i\Lambda({\bf r})}\sum_{i=1}^n\psi_i({\bf r})
\equiv\sum_{i=1}^n\Psi_i({\bf r})$. Each bosonic field $\phi_i$ is related to the projection
of $\Psi_i({\bf r})$ on the gapless edge modes \cite{compedge}. Denoting the projections by
$\Psi_i(x)$ one finds
\begin{equation}
\Psi_i(x) = \frac{F_i}{\sqrt{2\pi a}}e^{\frac{ 2\pi i}{L} N_i x+ i \phi_i(x)},
\label{operator}
\end{equation}
with corresponding normal ordered densities
\begin{equation}
:\!\rho_i(x)\!:\;=\;:\!\Psi_i(x)^\dagger\Psi_i(x)\!:\;=\sum_{j=1}^m K^{-1}_{ij}
\left(\frac{1}{2\pi}\partial_x\phi_j+\frac{1}{L}N_i\right).
\label{densities}
\end{equation}
Here $L$ is the length of the edge and $a$ is a short distance cutoff of the order of
the magnetic length. The Klein factors, $F_i$, and the conjugated number operators, $N_i$,
obey the algebra $\{ F_i,F_j \}=0$ for $i\neq j$,  $\{ F_i,F_j^\dagger
\}=2\delta_{ij}$ and $[F_i,N_j ]=F_i\delta_{ij}$. Finally, the statistics of the edge
channels and their interactions are given by the matrices
\begin{equation}
K= \left( \begin{array}{cc} 3 & 2 \\ 2 & 3 \end{array} \right),
~~~~~ V =\left( \begin{array}{cc} v & g \\ g & v
 \end{array} \right).
\label{KV}
\end{equation}
The matrix $V$ contains the
strength $g$ of the short-range interaction between the fermions while $v$ is the
velocity of the channels as determined by the confining potential and the interactions.
For simplicity we have assumed the same velocity for the two channels and ignored
the difference in their Fermi wavevectors which become irrelevant in the
strong interaction regime which we consider.

We derive an effective one-dimensional action describing the correlated tunneling
processes by integrating Eq. (\ref{interaction}) over the coordinate perpendicular to the edge
and replacing the composite fermion operators by their projections $\psi_i(x)$ on the edge
modes. Taking $\psi_i^\dagger(x)\psi_j(x)=\Psi_i^\dagger(x)\Psi_j(x)$ \cite{comment2},
and using Eq. (\ref{operator}), we find
\begin{eqnarray}
\nonumber
S_1&=& \int dx d\tau \left( \bar{\lambda}+ \lambda  \sum_{i=1}^2:\!\psi_i^\dagger\psi_i
\!:\right)\left( \psi_1^\dagger \psi_2 + {\rm h.c.} \right) \\
\nonumber
&=&\frac{1}{4\pi a}\int dx d\tau \left[ F_1^\dagger F_2 e^{\frac{ 2\pi i}{L} (N_2-N_1) x +
i\phi_2-i\phi_1} + {\rm h.c.} \right] \\
&\times& \left[ 2\bar{\lambda}+ \frac{\lambda}{5\pi}
\left( \partial_x \phi_1 +\partial_x \phi_2 + \frac{2\pi}{L}(N_1+N_2) \right)\right], 
\label{S1}
\end{eqnarray}
where we have used that $\sum_{i=1}^n K^{-1}_{ij}=\nu/n$ for a Jain fraction
$\nu=n/(2n\pm 1)$ \cite{compedge}. The strength of the correlated tunneling is
characterized by the coupling constant $\lambda$ and we have also included
a term, proportional to $\bar{\lambda}$, describing simple tunneling between
the levels, whose origin is the constant average value of the density in Eq.
(\ref{interaction}).

The action $S=S_0+S_1$ can be diagonalized exactly. To this end we add to it the
action of a free chiral auxiliary field, $\phi_0$,  with velocity $v_0=v-g$ \cite{Naud00}
\begin{eqnarray}
\nonumber
S_{\rm aux} &=& \frac{1}{4 \pi} \int dx d\tau \left[i \partial_x \phi_0 \partial_\tau \phi_0
+ v_0\left(\partial_x \phi_0\right)^2\right] \\
&+&\frac{\pi v_0}{L} \int d\tau  N_0^2 ,
\end{eqnarray}
and transform to new fields, $\varphi_i= \sum_j A_{ij} \phi_j$,
and number operators ${\cal  N}_i= \sum_j B_{ij} N_j$ \cite{comment3}, with
\begin{eqnarray}
\!\!A= \left( \begin{array}{ccc} \frac{1}{\sqrt{2}} & \frac{1}{2} & -\frac{1}{2} \\
\frac{1}{\sqrt{2}} & -\frac{1}{2} & \frac{1}{2} \\
0 & \frac{1}{\sqrt{10}} & \frac{1}{\sqrt{10}} \end{array} \right), ~~ B=
\frac{1}{2} \left( \begin{array}{ccc} 1 & 1 & -1 \\
1 & -1 & 1 \\
-1 & 1 & 1 \end{array} \right).
\end{eqnarray}
Defining new Klein factors ${\cal F}_0= F_0 F_1$,
${\cal F}_1=F_0 F_2$ and  ${\cal F}_2= F_1 F_2$, which are readily verified to
obey the required commutation algebra among themselves and with the ${\cal N}_i$,
allows us to introduce the fermions
\begin{equation}
\xi_i(x)=\frac{{\cal F}_i}{\sqrt{2 \pi a}} e^{ \frac{2 \pi i}{L} {\cal
    N}_i x + i \varphi_i(x)},~~~~~i=0,1
\label{newfermions}
\end{equation}
in terms of which the action $S=S_0+S_1+S_{\rm aux}$ reads
\begin{eqnarray}
\nonumber
S&=&\int dx d\tau\sum_{j=0,1}\left(\xi_j^\dagger\partial_\tau\xi_j-:\!iv_n
\xi_j^\dagger\partial_x\xi_j\!: \right) \\
\nonumber
&+&\frac{1}{4\pi}\int dx d\tau \left[ i\partial_x \varphi_2
\partial_\tau  \varphi_2+v_c (\partial_x \varphi_2)^2 \right] \\
\nonumber
&+&\frac{\pi}{L}\int d\tau \left[\frac{v_n}{2}\left({\cal N}_0+{\cal N}_1\right)^2+
v_c{\cal N}_c^2 \right] \\
\nonumber
&+&\int dx d\tau \left[\bar{\lambda}+\frac{\lambda}{\sqrt{10}\pi}\left(\partial_x
\varphi_2+\frac{2\pi}{L}{\cal N}_c\right)\right] \\
&\times&\left(\xi_0^\dagger\xi_1+\xi_1^\dagger\xi_0\right),
\label{transS1}
\end{eqnarray}
where ${\cal N}_c=({\cal N}_0+{\cal N}_1+2{\cal N}_2)/\sqrt{10}=(N_1+N_2)/\sqrt{10}$.
The fields $\xi_{0,1}$ which involve the neutral edge mode $\phi_1-\phi_2$ move with
its velocity $v_n=v-g$, while $\varphi_2$ which creates the symmetric combination
$\phi_1+\phi_2$ propagates with the velocity of the charged edge mode $v_c=5(v+g)$.

Next, we transform to the even and odd combinations of $\xi_{0,1}$
and bosonize once again
\begin{equation}
\xi_\pm = \frac{1}{\sqrt{2}}e^{\pm i \frac{\bar{\lambda}}{v_n}x}
(\xi_0 \pm \xi_1)  =\frac{{\cal F}_\pm }{\sqrt{2 \pi a}}
 e^{ \frac{2 \pi i}{L} {\cal  N}_\pm x + i \varphi_\pm(x)}.
\label{xipm}
\end{equation}
This step eliminates $\bar{\lambda}$ from $S$ and turns the correlated tunneling term into
a density-density interaction between $\varphi_2$ and $\varphi_{\pm}$. Finally, the resulting
quadratic action can be diagonalized by the transformation
\begin{equation}
\left( \begin{array}{c} \theta_0,Q_0 \\ \theta_1,Q_1 \\ \theta_2,Q_2 \end{array} \right)
=\left( \begin{array}{ccc} \frac{1}{\sqrt{2}} & \frac{1}{\sqrt{2}} & 0 \\
\frac{\cos \gamma}{\sqrt{2}} & -\frac{\cos \gamma}{\sqrt{2}} & -\sin \gamma \\
\frac{\sin \gamma}{\sqrt{2}} & -\frac{\sin \gamma}{\sqrt{2}} & \cos \gamma
 \end{array} \right)
\left( \begin{array}{c} \varphi_+,{\cal N}_+ \\ \varphi_-,{\cal N}_- \\
\varphi_2,{\cal N}_c \end{array} \right),
\label{finaltrans}
\end{equation}
where the rotation angle
\begin{equation}
\tan\gamma=\sqrt{(v_n-u_1)/(u_2-v_n)} \, ,
\label{gamma}
\end{equation}
is determined by the velocities
\begin{equation}
u_{1,2}= \frac{1}{2}\left[v_c + v_n\mp \sqrt{ \left(v_c-v_n\right)^2+ 4\lambda^2/5\pi^2}\right],
\label{velocities}
\end{equation}
of the physical modes in the diagonalized action
\begin{eqnarray}
S &=& \frac{1}{4\pi} \int dx d \tau  \sum_{j=0}^2 \left[ i
\partial_x \theta_j \partial_\tau \theta_j + u_j
\left( \partial_x \theta_j \right)^2 \right] \nonumber \\
&+& \frac{\pi}{L} \int d\tau \left( \sum_{j=0}^2 u_j Q_j^2+ u_0 Q_0^2\right) .
\label{Sfinal}
\end{eqnarray}
The $\theta_0$ mode is a new rotated auxiliary field moving with the same velocity
$u_0=v_0=v_n$ as the original one. As expected, and as shown below, its dynamics does
not enter into the electronic edge correlators.

Eqs. (\ref{velocities},\ref{Sfinal}) imply that the system becomes unstable when $u_1<0$,
and Eq. (\ref{finaltrans}) indicates that like in the toy model considered at the beginning
of this Letter, the instability involves a divergence of the total edge charge
$N_1+N_2\rightarrow\pm\infty$. However, in contrast to the toy model, the instability occurs
only once the correlated tunneling strength crosses the following threshold
\begin{equation}
\lambda > \sqrt{5}\pi\sqrt{v_c v_n}.
\label{condition}
\end{equation}
Composite fermions Hartree-Fock calculations of the collective modes of a $\nu=2/5$ edge
\cite{Nguyen04} find $v_c\sim v_n\sim 0.05 e^2/\epsilon$, where $\epsilon$ is the dielectric
constant. Since the correlated
tunneling coupling constant is $\lambda\sim e^2/\epsilon$, condition
(\ref{condition}) may well be satisfied.
\begin{figure}[t]
\includegraphics[width=7.0cm]{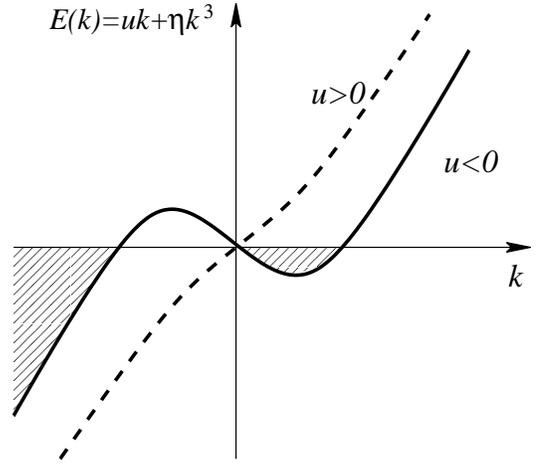}
\caption{The dispersion of chiral fermions described by the bosonized action (\ref{regular}).
When $u<0$, the fermions occupy the hashed regions and a droplet breaks from the main Fermi
sea.}
\label{edgefig}
\end{figure}

Evidently, the appearance of a negative velocity in the action (\ref{Sfinal})
calls for regularization. Higher order non-linear terms which control the
instability appear naturally once corrections to the assumed linear dispersion
of the edge composite fermion are taken into account in the CTL action (\ref{S0}).
In the course of the various transformations utilized above such terms generate
couplings between the modes as well as regularize any divergences. To focus the
discussion we ignore the former and consider the effect of a cubic term,
$\eta k^3$, in the bare edge dispersion, on the unstable mode, whose action is now
\begin{eqnarray}
\nonumber
&&\hspace{-0.3cm}S_{u}=\frac{1}{4\pi} \int  dx d\tau {\Bigg [}
i \partial_x \theta \partial_\tau \theta + u
\left(\partial_x\theta\right)^2  \\
&&\hspace{2.25cm}+\;\frac{\eta_2}{2}(\partial^2_x\theta)^2 +
\frac{\eta_4}{2}(\partial_x\theta)^4{\Bigg ]},
\label{regular}
\end{eqnarray}
where $\eta_2$ and $\eta_4$ are functions of $\eta$ and the other couplings
of our problem.
This action my be interpreted as describing chiral fermions with dispersion
$E(k)= u k+ \eta_4 k^3$ and residual self-interaction $(\eta_2-\eta_4)(\partial_x\rho)^2$.
When the velocity is positive, $u>0$, the nonlinear terms have little effect and the ground
state constitutes a Fermi sea that occupies the range $k<0$. However, if
$u<0$, the Fermi sea breaks into two distinct components, as illustrated in Fig. \ref{edgefig},
and the single chiral channel, which exists before the instability, turns into three.
Beside the mode which runs along the shifted edge of the original Fermi sea the
instability generates two additional counter-propagating channels on the edges of
the detached droplet. Note that the instability studied here is different from the one
discussed in Ref. \onlinecite{Yang03}, where the velocity remains positive for small $k$,
and the instability is caused by an interaction-induced negative
cubic term in the dispersion.

The effect of the instability on the edge local tunneling density of states (LDOS), as
determined by the electronic Green's functions
$G_j(x=0,\tau)= \langle\Psi_j(0,\tau)\Psi_j^\dagger(0,0)\rangle$, depends
on the nature of the correlations which exist between the electrons in the new droplet.
Deferring a discussion of this issue to a future publication \cite{longpaper} we would
like to elucidate here the role of correlated tunneling in the LDOS before the instability
sets it. Using the fact that the auxiliary field $\phi_0$ is completely decoupled from
the physical fields, and Eq. (\ref{newfermions}), we find
\begin{eqnarray}
\nonumber
G_j(x,\tau)&=&\frac{\langle\xi_{j-1}(x,\tau)\xi_{j-1}^\dagger(0,0)
e^{i\sqrt{\frac{5}{2}}[\varphi_2(x,\tau)-\varphi_2(0,0)]}\rangle}
{\langle e^{\frac{i}{\sqrt{2}}[\phi_0(x,\tau)-\phi_0(0,0)]}\rangle} \\
&\propto& \sum_{\sigma=\pm}e^{i\sigma\frac{\bar{\lambda}}{v_n}x}
\prod_{j=1}^2(x+iu_j\tau)^{-\alpha_{j,\sigma}},
\label{LDOS}
\end{eqnarray}
where
\begin{equation}
\alpha_{j,\sigma}=\left[3+(-1)^j\left(2\cos 2\gamma +
\sqrt{5}\sigma\sin 2\gamma\right)\right]/2 .
\label{alpha12}
\end{equation}
Note that the dynamics of the auxiliary field does not appear in the correlators, which
reduce in the limit $\lambda\rightarrow 0$ to the known CTL result \cite{Naud00}.
Moreover, before the instability $G_j(x=0,\tau)\sim\tau^{-3}$;
the same power-law predicted by the CTL model for the LDOS of a $\nu=2/5$ edge.

We have extended the above analysis to two additional fractions,
$\nu=2/3$ and $\nu=3/7$ \cite{longpaper} and found a similar instability in
both cases. Using an analogous action to the one given by Eqs. (\ref{S0},\ref{S1})
one obtains that the condition for the instability in the $\nu=2/3$ case is
$\lambda>\sqrt{3}\pi\sqrt{v_c v_n}$, where $v_c= 3(v+g)$ and $v_n= v-g$.
It turns out that since the charge and neutral modes of the $\nu=2/3$ edge propagate
in opposite directions, the correlated tunneling processes tend to reduce the tunneling
exponent from its CTL value of $\alpha=2$, even before the instability occurs.
At filling fraction $\nu=3/7$, the edge, which supports one charged mode, with velocity
$v_c=7(v+2g)$, and two neutral modes, with velocity $v_n=v-g$, reconstructs
once $\lambda>\sqrt{7/18}\pi\sqrt{v_cv_n}$. These results suggest a possible
breakdown of the naive CTL model for the FQH edge in the entire range
$1/3<\nu<1$.

We thank Nathan Andrei, Eldad Bettelheim, and Paul Wiegmann for
useful discussions. This research was supported by the United States
- Israel Binational Science Foundation (grant Nos. 2004162 and
2004128), and by the Israel Science Foundation (ISF) funded by the
Israeli Academy of Science and Humanities (grant No. 63/05).


\begin{thebibliography}{99}
\bibitem{comment3} The original number operator, $N_i$ have an integral spectrum. The
operators ${\cal N}_i$ have an integral spectrum or an integral spectrum shifted by $1/2$.
However, since the tunneling term changes ${\cal N}_0$ and ${\cal N}_1$ by $\pm 1$,
the Hilbert space breaks into sectors which remain disconnected over time.
\end{thebibliography}
\end{document}